\documentclass[conference]{IEEEtran}
\IEEEoverridecommandlockouts
\usepackage{cite}
\usepackage{caption}
\usepackage{amsmath,amssymb,amsfonts}
\usepackage{algorithm}
\usepackage{algpseudocode}
\usepackage{enumitem}
\usepackage{balance}
\usepackage{graphicx}
\usepackage{textcomp}
\usepackage{threeparttable}
\usepackage{amsmath}
\usepackage{comment}
\usepackage{xcolor}
\usepackage{xurl}
\usepackage[breaklinks]{hyperref}
\def\BibTeX{{\rm B\kern-.05em{\sc i\kern-.025em b}\kern-.08em
    T\kern-.1667em\lower.7ex\hbox{E}\kern-.125emX}}

\begin{document}

\title{Automated Validation of Insurance Applications against Calculation Specifications}
\author{\IEEEauthorblockN{Advaita Datar\textsuperscript{$\ast$}}
\IEEEauthorblockA{\textit{Tata Research Development and} \\
\textit{Design Centre (TRDDC), TCS Research,}\\
Pune, India \\
advaita.datar@tcs.com}
\and
\IEEEauthorblockN{Amey Zare\textsuperscript{$\ast$} \thanks{\textsuperscript{$\ast$} Equal Contribution}}
\IEEEauthorblockA{\textit{Tata Research Development and} \\
\textit{Design Centre (TRDDC), TCS Research,}\\
Pune, India \\ 
amey.zare@tcs.com}
\and
\IEEEauthorblockN{Asia A}
\IEEEauthorblockA{\textit{Tata Research Development and} \\
\textit{Design Centre (TRDDC), TCS Research,}\\
Pune, India \\
asia.4@tcs.com}
\and
\IEEEauthorblockN{R Venkatesh}
\IEEEauthorblockA{\textit{Tata Research Development and} \\
\textit{Design Centre (TRDDC), TCS Research,}\\
Pune, India \\
r.venky@tcs.com}
\and
\IEEEauthorblockN{Dr. Shrawan Kumar}
\IEEEauthorblockA{\textit{Tata Research Development and} \\
\textit{Design Centre (TRDDC), TCS Research,}\\
Pune, India \\
shrawan.kumar@tcs.com}
\and
\IEEEauthorblockN{Ulka Shrotri}
\IEEEauthorblockA{\textit{Tata Research Development and} \\
\textit{Design Centre (TRDDC), TCS Research,}\\
Pune, India \\
ulka.s@tcs.com}
}

\maketitle
\begin{abstract}
Insurance companies rely on their Legacy Insurance System (LIS) to govern day-to-day operations. These LIS operate as per the company’s business rules that are formally specified in Calculation Specification (CS) sheets. To meet ever-changing business demands, insurance companies are increasingly transforming their outdated LIS to modern Policy Administration Systems (PAS). Quality Assurance (QA) of such PAS involves manual validation of calculations’ implementation against the corresponding CS sheets from the LIS. This manual QA approach is effort-intensive and error-prone, which may fail to detect inconsistencies in PAS implementations and ultimately result in monetary loss. To address this challenge, we propose a novel low-code/no-code technique to automatically validate PAS implementation against CS sheets. Our technique has been evaluated on a digital transformation project of a large insurance company on 12 real-world calculations through 254 policies. The evaluation resulted in effort savings of approximately 92 percent against the conventional manual validation approach.
\end{abstract}

\section{Introduction}
\begin{figure*}[t]
\centering
\fbox{\includegraphics[width=0.98\textwidth]{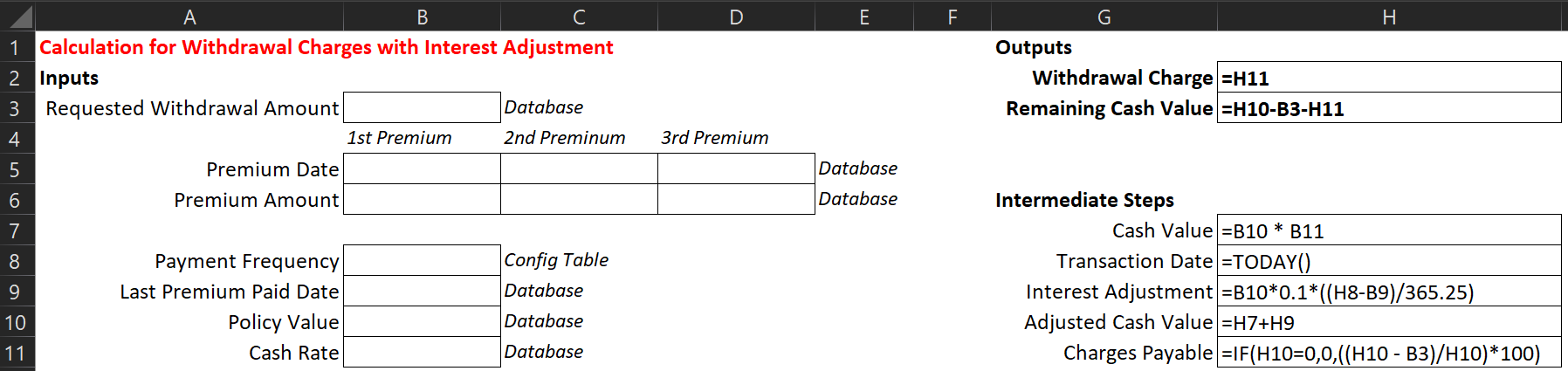}}
\caption{A Sample Calculation Specification (CS) Sheet that Specifies Business Rules for Withdrawal Charge Calculation}
\label{fig:CS}
\end{figure*}

Insurance companies conduct business with their customers through Legacy Insurance Systems (LIS) that operate as per the company's business rules. These business rules are essential to maintain data consistency in the company's LIS and to avoid monetary loss. In LIS, these business rules are implemented in the form of financial calculations that are formally specified in spreadsheets that are known as Calculation Specification (CS) sheets (Fig. \ref{fig:CS}). These CS sheets are prepared by the domain experts of the insurance company, and they mimic the functioning of the LIS for a given calculation. To meet the high demands and extremely dynamic nature of modern-day insurance industry, LIS need to be digitally transformed to modern Policy Administration Systems (PAS). Nowadays, digital transformation is necessary for the survival of insurance companies and hence they are making substantial investments for the same \cite{gartnerAutomation}. The motivation for our work comes from the struggles of one such digital transformation initiative for the LIS of a large insurance company.

When LIS are transformed to PAS, insurance companies perform Quality Assurance (QA) of PAS on their entire book of business to ensure that PAS operates as per expectations. The conventional approach for QA involves manual validation of PAS implementation against the business rules implemented in the LIS. In general, LIS are large, complex and not accessible to third party QA team, which makes QA of PAS against LIS a big challenge \cite{oracleInsurance,tenjinYethi}. As an alternative, CS sheets, that formally specify the company's business rules, are used for performing QA of PAS. In a typical CS sheet (Fig. \ref{fig:CS}), the entered input values (cells B3-B11) trigger intermediate calculation steps (cells H7-H11) to generate the expected output values (cells H2-H3). These expected output values need to be compared with the output values generated by PAS to ensure consistency between modern PAS and LIS. 
\begin{figure*}[t]
\centering
\fbox{\includegraphics[width=0.98\textwidth]{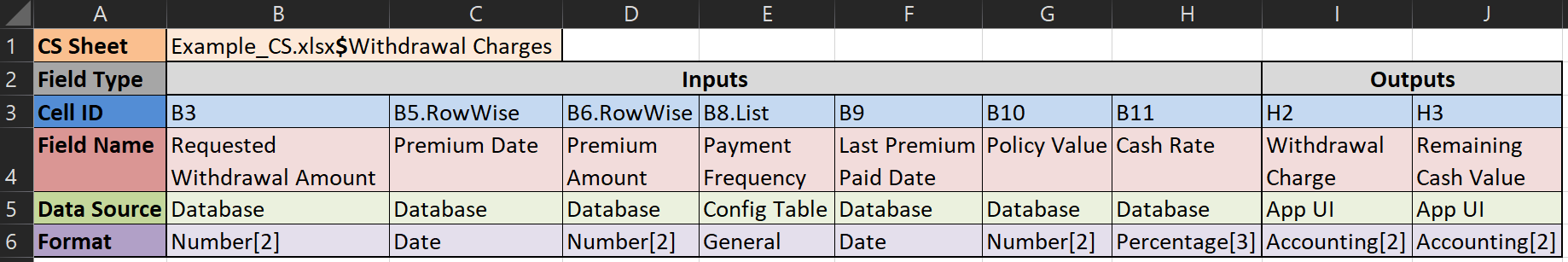}}
\caption{Schema Corresponding to the Calculation Specification (CS) Sheet in Fig. \ref{fig:CS}}
\label{fig:Schema}
\end{figure*}

The validation of modern PAS against CS sheets is typically done manually by considering real-world policies as test data. For this, QA team \cite{manualValidation} manually i) collects the required input and output values for a given policy from respective data sources (cells C3, E5 etc. in Fig. \ref{fig:CS}), ii) enters the input values in input cells of CS sheet, and iii) compares the expected output values generated by CS sheet with the output values generated in PAS for the same set of input values. Such a manual QA approach is tedious, effort-intensive and a big challenge, especially in real-world scenarios when \emph{millions of policies} need to be validated against \emph{thousands of CS sheets}. Hence, there is a clear need for a robust and flexible technique for automated validation of calculation implementation (PAS) against its specification (CS sheets).

To address this need, we propose a technique that automates the aforementioned manual QA approach. This technique proposes a novel metadata structure for CS sheet, referred henceforth as \emph{schema}, to capture the input and output details required for validation. Furthermore, we have developed a mechanism for collecting input and output data by referring schema file and the details provided by domain experts. The inputs are entered in the CS sheet, and the output generated by the CS sheet is compared with the output collected from PAS. Thus, the proposed technique is specification-agnostic that enables an efficient and reliable validation of PAS implementation against corresponding calculation specification. We have implemented our technique in the form of a TCS \cite{tcsRef} internal tool: Batch-wise Validation of Code against Specification (BVCS), which
\begin{itemize}
\item enables automated validation of millions of policies against a wide variety of CS sheets.
\item is adaptable to any calculation specification described in spreadsheet format.
\item is a low-code/no-code tool that does not need changes in its code in case there are changes in the input CS sheet.
\end{itemize}

The rest of the paper is organized as follows. Section \ref{approach} describes BVCS in detail. Section \ref{experiments} presents BVCS evaluation results. In Section \ref{related} we discuss literature related to BVCS, and we conclude in Section \ref{conclusion}.
\section{Proposed Solution}\label{approach}
\begin{figure}[t]
    \centering
    \fbox{\includegraphics[width=0.95\columnwidth]{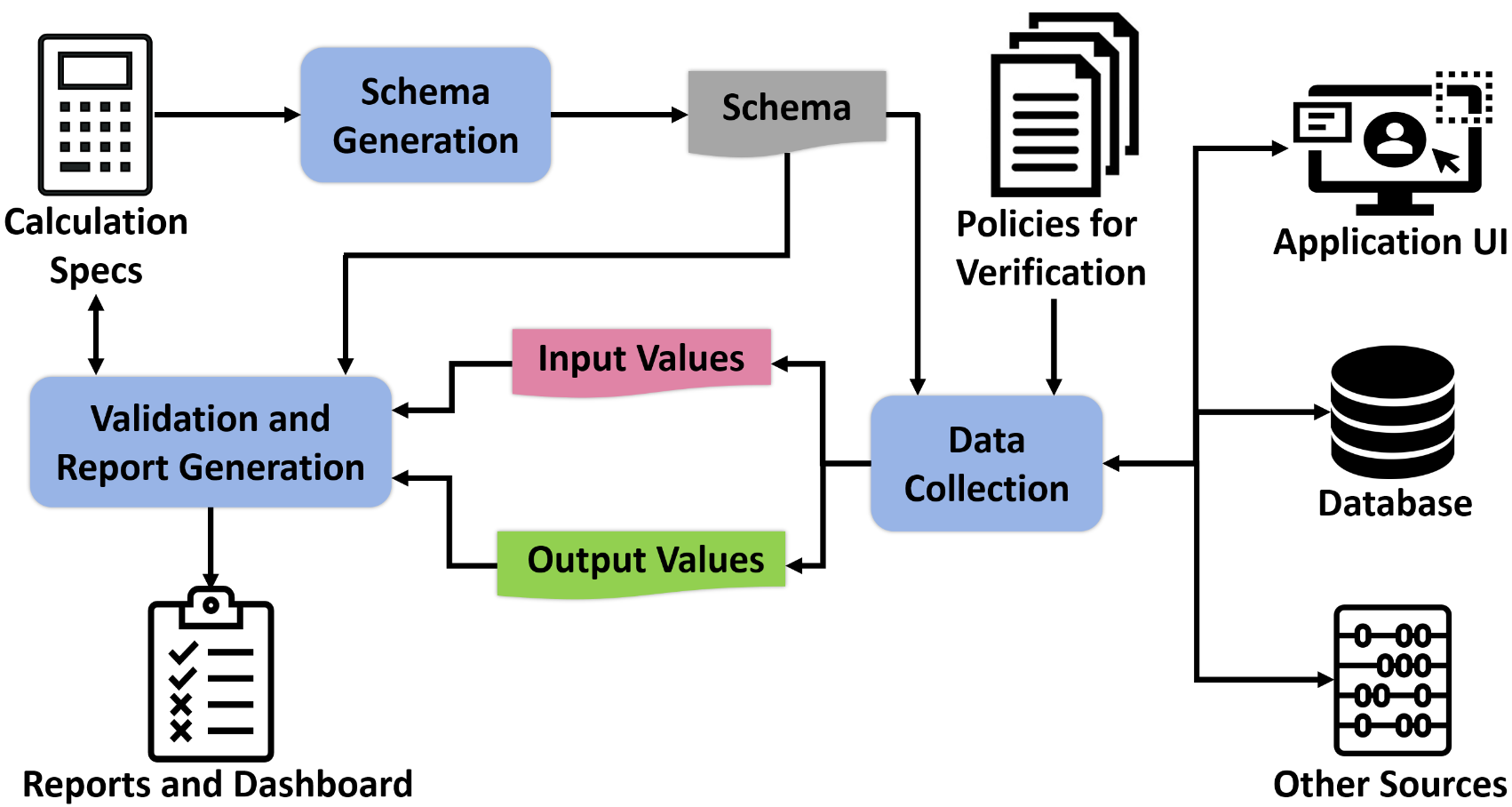}}
    \caption{Workflow of Proposed Approach}
    \label{fig:Approach}
\end{figure}

This section describes our technique Batch-wise Validation of Code against Specification (BVCS) using the CS sheet in Fig. \ref{fig:CS}. BVCS is a 3-step technique that automatically
\begin{enumerate}
\item generates a metadata file, referred henceforth as \emph{schema}, corresponding to the CS sheet,
\item collects input-output values, from respective data sources of PAS, that are required for validation,
\item fills input values into the CS sheet and compares expected output values generated by the CS sheet, with the output values generate by PAS, and shows the validation results through an interactive dashboard that highlights output value mismatches.
\end{enumerate}
The high-level overview of BVCS is shown in Fig. \ref{fig:Approach}. The key steps and features of BVCS are described as follows.
\begin{algorithm}
\caption{Generate Schema for CS} \label{alg:schemaGen}
\begin{algorithmic}[1]
\Procedure {generateSchema}{$C$}
\State $ queue \leftarrow [..] $
\State push $CS$ into $queue$  
\While{$queue$ is not empty}
    \State pop $CS$ from $queue$
    \State $In,Out,RS \gets getInputOutputCells(CS)$
    \State $Sheets \gets addSheet({CS,\{In,Out\}})$
    \For{$rs \in RS$}
        \If{ $! processed($rs$)$ }
           \State push $rs$ into $queue$ 
        \EndIf
    \EndFor
\EndWhile
\State \Return{$Sheets$}
\EndProcedure
\Statex
\Procedure {getInputOutputCells}{$CS$}
    \State $Cells \gets getAllUsedCells($CS$)$
\For {$c \in Cells $}
\If{ $! nodeAvail(Nodes,c)$}
     \State $ Nodes\gets createNode(c)$
\EndIf
\If{$ getInputOutputCells(c)$}
    \State $ RC, RS \gets parseFormula(c) $
    \State \textcolor{blue}{//\textit{RC-referred cells, RS-referred sheets}}
    \For {$d \in RC $}
        \If{ $! nodeAvail(Nodes, d)$ }
           \State $ Nodes\gets createNode(d)$
        \EndIf
     \State $Edges \gets drawEdge(c,d)$   
      \State  \textcolor{blue}{// \textit{c-source node, d-destination node}}
    \EndFor
\EndIf
\EndFor
 \For {$c \in Cells $}
  \If{$ ! src(Edges,c)$ \&\& $des(Edges,c)$ }    
      \State $ InputCells \gets addCell(c) $
   \ElsIf{$ ! des(Edges,c)$ \&\& $src(Edges,c)$ }
     \State $ OutputCells \gets addCell(c) $
    \EndIf
\EndFor
\State \Return{$ InputCells, OutputCells, RS$}
\EndProcedure
\end{algorithmic}
\end{algorithm}

\subsection*{Step 1: Schema Generation}\label{schemaGen}
Fig. \ref{fig:Schema} shows the schema corresponding to the example CS sheet in Fig. \ref{fig:CS}. Schema is a representation of the metadata of CS sheet, and it contains the input and output details required for validation of PAS against the corresponding CS sheet. In order to devise a robust schema generation technique, we analyzed various CS sheets and generalized the characteristics of input and output cells. This was done to ensure that BVCS generates schema for CS sheets that do not strictly adhere to their formatting guidelines. Based on our observations, we define input and output cells as follows.
\begin{itemize}
\item \emph{Input Cell:} A cell that is not dependent\cite{cellReference} on any other cell for it's value, but is referred by other cell(s). Examples, cells B3, B5, C5 etc. in Fig. \ref{fig:CS}.
\item \emph{Output Cell:} A cell that is not referred by any other cell, but it's value is calculated based contents of other cell(s). Examples, cells H2 and H3 in Fig. \ref{fig:CS}.
\end{itemize}
 
The algorithm used by BVCS for schema generation is shown in Algorithm \ref{alg:schemaGen}. To generate schema for a given CS sheet, BVCS firstly reads the contents of all the cells, that aren't empty, to identify \emph{referred cells}. These are identified by parsing the formula in each nonempty cell. For example, in Fig. \ref{fig:CS} cell H3 has a formula, ``=H10-B3-H11", and hence H10, B3 and H11 are identified as referred cells for H3. In case a cell of the input CS sheet has referred cells belonging to a different CS sheet, then that sheet is tagged as \emph{referred sheet}. It is worth noting that the example CS sheet in Fig. \ref{fig:CS} doesn't have any referred sheets. The identified referred cells and referred sheets drive the entire process to find input and output cells, which is described as follows.

BVCS maintains the relationship between a cell and it's \emph{referred cell(s)} in the form of a directed graph. Each cell (e.g. H3 in Fig. \ref{fig:CS}) is tagged as `destination node' and an edge is drawn from each of it's referred cells (H10, B3, H11 in Fig. \ref{fig:CS}) by tagging them as `source nodes'. Next, the edges between each cell and corresponding referred cells are traversed transitively to identify input and output cells as per their aforementioned definition. Input cells are the ones that are never tagged as a `destination node' but are tagged as `source node' at least once. Output cells are the ones that are never tagged as a `source node' but are tagged as `destination node' at least once.

Once all input and output cells are identified, the BVCS captures their respective a) data sources mentioned in the CS sheet (cells C3, E5 etc. in Fig. \ref{fig:CS}), and b) value format. Lastly, BVCS dumps all collected information in a Comma-Separated Value (CSV) file as shown in \ref{fig:Schema}. The resulting schema file is a lightweight alternative to the input CS sheet and acts as a foundation for gathering data for validating PAS against the input CS sheet. The contents of each row of the schema are described below.
\begin{itemize}
\item \emph{CS Sheet} is the spreadsheet file name and tab name, separated by \emph{\$}, that contains calculation specifications.
\item \emph{Field Type} indicates whether given cell is input cell or output cell.
\item \emph{Cell ID} in which input values are to be entered or output values need to be read from, depending on whether the cell is an input or output cell. Notice that cell IDs for cells that are part of a table (Premium Date and Premium Amount in Fig. \ref{fig:CS}) are appended with \emph{RowWise} and \emph{ColumnWise} depending on the direction in which the respective table expands.
\item \emph{Data Source} from where data for input and output values is to be retrieved. Note that output values are always retrieved from the user interface (UI) of PAS.
\item \emph{Format} of the values to be entered/read, as per the formats supported in Microsoft Excel. Square brackets indicate the decimal precision of numeric formats.
\end{itemize}
\subsection*{Step 2: Data Collection}\label{collect}
Once the schema is ready, BVCS collects the input and output values for the policy under test from the PAS, based on the sources mentioned in the \emph{Data Source} row of the schema. To retrieve data from \emph{App UI} of PAS, we use Selenium scripts\cite{seleniumT}, and to retrieve data from \emph{Database} and \emph{Config Table} we use Python scripts. These data collection scripts are designed to collect data for any given CS sheet based on parameters like UI screen name, field name, database queries etc. These CS-specific parameters need to be provided by domain experts for each CS sheet under test.

For example, to collect value(s) for `Policy Value' (cell B10 in Fig. \ref{fig:CS}), the domain expert is expected to provide the respective parameterized SQL query\cite{parSQ}. Similarly, the domain expert needs to provide the name of UI screen/page that contains the value of `Withdrawal Charge' (cell H2 in Fig. \ref{fig:CS}). Based on the provided information, respective Selenium and Python scripts interact with the database and application UI to retrieve input and output values for the policy under test. Collection and execution of input and output values, based on such parameterized information, makes BVCS a low code/no code test automation technique. Therefore, to validate a different input CS sheet, BVCS does not require change in its code. It only needs the parameterized information for collecting data for the different CS sheet.
\subsection*{Step 3: Validation and Report Generation}
Once all the input-output values for the policy under test have been retrieved from PAS, BVCS validates the collected data against corresponding CS sheet and it's referred sheets' data points, and reports the results. BVCS performs validation by using the CS sheet as a black-box for entering inputs and checking outputs. First, all input values are entered in respective cell IDs, which triggers the intermediate steps that generate expected output values in the respective cell IDs of the CS sheet. In case referred sheets have been identified for the input CS sheet, there is no restriction on the order for entering inputs in the input CS sheet and referred sheets. As long as data is entered in all involved CS sheets, the generated output will always be considered valid.

Next, BVCS compares the expected output generated in CS sheet with the PAS output collected in Step \ref{collect}. The results of comparison are reported to the user via a dashboard that indicates whether the validation for the policy under test has \emph{PASSED} or \emph{FAILED}. Additionally, a copy of the filled CS sheet, that contains values for the policy under test, is preserved as an evidence of validation. In this evidence CS sheet, BVCS also highlights the cell(s) for which the value in CS sheet was found to be different from the value in PAS. This helps the BVCS user to quickly identify the data points where validation failed.
\subsection*{Key Feature: Batch-Wise Execution}\label{batch}
The schema is a lightweight alternative to the CS sheet as it contains textual facts about the CS sheet, and omits complex spreadsheet features and formulae. In fact, generation of schema and using it for validating PAS is the key innovative idea in BVCS. Schema generalizes the structure of calculation specifications and as a result domain experts could write parameterized data collection scripts that gave BVCS the flexibility to validate multiple CS sheets through multiple policies. The schema and these parameterized scripts make BVCS a low-code/no-code technique as they enable BVCS to validate different CS sheets without requiring any changes in BVCS code. This fact is exploited in the batch-wise execution mode of BVCS. In this mode, BVCS validates PAS against multiple input CS sheets through corresponding policies in bulk.

In order to properly validate PAS implementation against CS sheets, QA team needs to ensure that all statements, branches and possible execution paths in the PAS implementation are tested. In real-world scenario, this is achieved by using data of policies, relevant to the CS sheet, as test cases for QA. This ensures QA is done on actual test data, instead of randomly generated test data, and also ensures coverage of all data points for validation. Moreover, validating the data of all policies on all relevant CS sheets is necessary for ensuring test coverage and regulatory compliance of PAS. To achieve this, QA team can run BVCS in the batch-wise execution mode to validate multiple policies on multiple CS sheets by providing the list of policies and CS sheets as additional inputs. BVCS batch-wise execution is extremely useful in real-world scenarios where PAS needs to be validated against large number of CS sheets through bulk actual customer policies, as described in detail in Section \ref{experiments}.
\section{Experimental Results}\label{experiments}
BVCS was evaluated by a TCS \cite{tcsRef} QA team that was handling a digital transformation project for a large insurance company. The QA team used BVCS to validate the insurance company's PAS against 12 CS sheets through 254 real-world insurance policies. The domain experts from the insurance company provided these 12 CS sheets as representatives of the wide variety of CS sheets used by the company. These CS sheets contained specification of calculations for surrender charge, service charge, guaranteed minimum death benefit etc. The selected CS sheets belonged to the categories described in Table \ref{categoryTable}. The insurance policies under test were extracted from the company's production database and contained actual details of the policies purchased by the company's customers. The experiments were conducted on i7-8565U CPU @1.80GHz with 16GB RAM and Microsoft Windows 10 Pro OS. The objectives of the experiments were
\begin{enumerate}[label=\alph*)]
\item calculating effort savings through one-to-one comparison with manual approach, and
\item measuring performance of BVCS in batch-mode execution.
\end{enumerate}
\begin{table}[b]
\centering
\caption{Categorization of CS Sheets}
\begin{tabular}{|l|c|c|c|} 
\hline
Attribute of CS Sheet & \multicolumn{3}{|c|}{CS Sheet Category}\\
\cline{2-4}
 & Easy & Medium & Hard\\
\hline
Number of dependent sheets & 0  & 4  & 6 \\
\hline
Number of input fields & $<8$ & $8-15$ & $>15$\\
\hline
Any advance features? & No & No & Tabular Data,\\
 & & & config tables, etc.\\
\hline
\end{tabular}
\label{categoryTable}
\end{table}
\subsection{Comparison with Manual Approach for Single Policy}
To evaluate the efforts saved through BVCS, the efforts spent by a tester for end-to-end manual validation of a single insurance policy were compared against the time taken by BVCS to do the same. End-to-end validation comprises of collecting input \& output values from respective data source(s) of PAS, filling input values in the CS sheet, and comparing the output values generated by CS sheet with output values collected from PAS. 
\begin{table}[t]
\centering
\caption{Efforts comparison for a single policy}
\begin{tabular}{|c|c|c|c|c|c|c|}
 \hline
 CS & \multicolumn{3}{|c|}{BVCS Time (sec)} & \multicolumn{2}{|c|}{Total Efforts (min)} & Effort\\
 \cline{2-6}
 Sheet&\textit{Step 1}&\textit{Step 2}&\textit{Step 3}&BVCS &Manual & Savings(\%) \\
 \hline 
 Easy-1 & 1.3 & 157 & 4.5 & 2.71 & 20 & 86.43 \\
 \hline
 Easy-2 	& 3	& 110 &	17 & 2.17 &	20 & 89.17 \\
 \hline
 Easy-3 & 3 & 123 &	14 & 2.33 &	25 & 90.67\\
 \hline
 Med-1 &	3.3 & 129 &	13.6 & 2.43 & 30 & 91.89 \\
 \hline
 Med-2 &	2.2 & 143 &	7.9 & 2.55 & 30 & 91.49 \\
 \hline
 Med-3 &	1.8 & 185 &	12.08 &	3.31 & 40 &	91.71 \\
 \hline
 Med-4 &	3 &	180 & 15 &	3.30 &	40 & 91.75 \\
 \hline
 Med-5 &	3 &	160 & 14 & 2.95 & 40 &	92.63 \\
 \hline
 Med-6 & 3 &	170 & 16 & 3.15 & 40 &	92.13 \\
 \hline
 Hard-1 & 2.8 & 210 & 9.3 & 3.67 &	60 & 93.83 \\
 \hline
 Hard-2 & 3& 268 & 12.42 & 4.72 & 70 & 93.25 \\
 \hline
 Hard-3 & 3.3 &	283 & 15.6 & 5.03 &	75 & 93.29 \\
 \hline
\end{tabular}
\begin{tablenotes}[para,flushleft]
\centering [\textit{Step 1} - Schema Generation, \textit{Step 2} - Data Collection,\\\textit{Step 3} - Validation and Report Generation] 
\end{tablenotes}
\label{singleTable}
\end{table}

Table \ref{singleTable} shows the results of one-to-one comparison of BVCS against manual validation approach. In this experiment, BVCS outperformed manual approach with an average effort savings of approximately 92\%. It is worth noting that the efforts shown in Table \ref{singleTable} do not include the efforts spent by domain experts to provide CS-specific parameters, such as screen name, database query etc., for data collection scripts (refer Step \ref{collect}). This is because both manual and BVCS-based approach depend on the said CS-specific input parameters for end-to-end validation.
\subsection{Batch-Mode Execution for Multiple Policies}
To ensure that BVCS is robust and scales on bulk real-world data, 254 insurance policies relevant to CS sheets \emph{Easy-3} and \emph{Med-3} were validated end-to-end using BVCS. These two CS sheets were picked because they were representative of the common types of CS sheets and also had the maximum number of actual policies readily available.
\begin{table}[b]
\centering
\caption{BVCS batch-mode execution results}
\begin{tabular}{|c|c|c|c|c|c|} 
 \hline
 CS & No. of & BVCS & \multicolumn{2}{|c|}{Manual (mins)} & Effort\\
 \cline{4-5}
 Sheet & Policies & Effort & Per & Total & Savings\\
  &  & (mins)  & Policy & Effort & (\%)\\
 \hline
 Easy-3 & 171 & 360 & 25 & 4275 & 91.58\\
 \hline
 Med-3 & 83 & 350 & 40 & 3320 & 89.46\\
 \hline
\end{tabular}
\label{batchTable}
\end{table}

Table \ref{batchTable} shows observations from BVCS batch-mode execution and the comparison of BVCS machine efforts with projected efforts in case a tester manually validates the same number of policies. In this experiment, BVCS was able to scale-up on the real-world policies with effort savings of approximately 91\%. It is worth noting that,
\begin{itemize}
\item BVCS execution time (column \emph{`BVCS Effort'}) shows the time taken for data collection, validation and report generation only, as schema generation was effectively a one-time activity for execution of each CS sheet in batch mode.
\item The indicated manual effort is a ball-park estimate. It has been extrapolated by multiplying number of policies with the average manual effort spent by a tester to validate a single policy for respective CS sheet category.
\end{itemize}

\subsection*{Key Observations and Impact}
As evident from Table \ref{singleTable} and \ref{batchTable}, BVCS is very effective in automatically validating implementation of calculations against their specifications through bulk real-world data. BVCS is also very efficient and, on an average, results in approximately 92\% effort savings when compared to existing manual validation approach. Furthermore, the inconsistencies reported by BVCS, for the 12 CS sheets under test, matched the ones reported through manual validation. This makes BVCS a robust and reliable approach for QA of PAS.

The success of BVCS allows TCS QA team to plan acceptance testing of the PAS of a large insurance company on it's entire book of business. Use of BVCS will enable validating the company's PAS against thousands of CS sheets using millions of policies. Based on the results of conducted experiment, the insurance company estimates that use of BVCS in all their projects will lead to effort savings of 100 person years and cost savings of US \$10 million.
\section{Related Work}\label{related}
Quality assurance (QA) of digitally transformed PAS continues to be a challenge for insurance applications. In general, testers are not familiar with the insurance domain and underlying business rules and workflows, and hence domain experts assist in QA of PAS through rigorous manual review \cite{manualValidation}. On the other hand, techniques for QA through test automation \cite{StudyTestAutomation,tenjinYethi,TestAutoBFSI} typically test the implementation against test cases that do not necessarily check code compliance against calculation specification.

Despite recommendations from domain experts \cite{actuarialGuidelines}, very limited work has been done to validate implementation against calculation specifications. In some techniques \cite{calcToCode}, domain experts create a model corresponding to the calculation specifications, via a custom domain-specific language. Later this model is used to automatically generate code for the calculation. Few approaches also use the created model to automatically generate functional test cases \cite{restAPI}. However, all such model-based techniques completely rely on domain experts to review the models for correctness, and require additional efforts to modify the model for accommodating changes in specification.

Other approaches relevant to digital transformation use trace logging to validate the implementation of modern PAS against legacy application \cite{wordToMPS} by annotating key segments of both versions of code and running both applications on a regression test suite. Later, inconsistencies are identified by comparing the text in execution traces of modern and legacy applications. However, such approaches lead to a large number of false positives, and cannot be used when legacy code is not accessible due to confidentiality.
\section{Conclusion and Future Work}\label{conclusion}
In this paper, we have presented a technique \emph{BVCS} for validating calculation implementation against formal specification. It automatically validates code against specification spreadsheets i.e. CS sheets and presents the results to the user through an interactive dashboard. BVCS is a simple, resilient and low-code/no-code technique that can be extended to support other types of formal specification and other financial applications, such as banking, mutual funds etc., with minimal or no change in the BVCS implementation.

BVCS was evaluated on the PAS of a large insurance company using real-world policies, and was found to be reliable, robust and efficient with average effort savings of approximately 92\% against conventional manual approach. In the future, we plan to use BVCS to validate the PAS of the insurance company for their entire book of business (11 million policies) through 1200 CS sheets.
\section{Acknowledgments}
We would like to thank Punyakoti Sathish, Suresh Bhaskaramurthy and Kannan Dhanasekaran from TCS insurance business team for their assistance in formulating the problem statement and conducting experiments.
\balance
\bibliographystyle{ieeetr}
\bibliography{main}

%

\end{document}